\documentclass[a4paper]{article}%
\usepackage{amsmath}

\usepackage{amssymb}
\usepackage{amsfonts}
\usepackage[T1]{fontenc}
\usepackage{graphicx}%
\usepackage{epsfig}%
\usepackage{calc}%
\usepackage{dsfont}%Pour \mathds{R} par exemple
\usepackage{longtable}%
\usepackage{overcite}%
\usepackage{lscape}%
\usepackage{multirow}%

\newenvironment{aleq*}{\begin{equation*}\begin{aligned}}{\end{aligned}\end{equation*}}

\newenvironment{gaeq*}{\begin{equation*}\begin{gathered}}{\end{gathered}\end{equation*}}

\makeatletter
\renewcommand{\@biblabel}[1]{$^{#1}$}%
\makeatother%
\begin{document}%
\title{\bf Soliton and similarity solutions of $\mathcal{N}=2,4$ supersymmetric equations.}
\author{Laurent Delisle\thanks{email address: delisle@dms.umontreal.ca}, V\'eronique Hussin\thanks{email address: hussin@dms.umontreal.ca }\\
D\'epartement de math\'ematiques et de statistique, Universit\'e de Montr\'eal,\\
C.P. 6128, Succc. Centre-ville, Montr\'eal, (QC) H3C 3J7, Canada}%
\date{}
\maketitle%
\fontsize{12}{14} \selectfont
\begin{abstract}
We produce soliton and similarity solutions of supersymmetric extensions of Burgers, Korteweg-de Vries  and modified KdV equations. We give new representations of the $\tau$-functions in Hirota bilinear formalism. Chiral superfields are used to obtain such solutions. We also introduce new solitons called virtual solitons whose nonlinear interactions produce no phase shifts. 
\end{abstract}
{Running Title: Soliton and similarity solutions of $\mathcal{N}=2,4$ supersymmetric equations.\\
2010 Mathematics Subject Classification: 35Q51, 35Q53.\\
Keywords: Supersymmetric equations, solitons and Hirota bilinear formalism.}
\section{Introduction.}
 The study of $\mathcal{N}=2$ supersymmetric (SUSY) extensions of nonlinear evolution equations has been largely studied in the past \cite{LM,AHW,IMM,GS,HK,ZP,DH,ZLW} in terms of integrability conditions and solutions. Such extensions are given as a Grassmann-valued partial differential equation with one dependent variable $A(x,t;\theta_1,\theta_2)$ which is assumed to be bosonic to get non trivial extensions. 
The independent variables are given as a set of even (commuting) space $x$ and time $t$ variables and a set of odd (anticommuting) variables $\theta_1$, $\theta_2$.
 Since the odd variables satisfy $\theta_1^2=\theta_2^2=\{\theta_1,\theta_2\}=0$, the dependent variable $A$ admits the following finite Taylor expansion
\begin{equation}
 A(x,t;\theta_1,\theta_2)=u(x,t)+\theta_1\xi_1(x,t)+\theta_2\xi_2(x,t)+\theta_1\theta_2v(x,t),\label{fieldA}
\end{equation}
where $u$ and $v$ are bosonic complex valued functions and $\xi_1$ and $\xi_2$ are fermionic complex valued functions. 
In this paper, we show that some of these extensions can be related to a linear partial differential equation (PDE) by assuming that $A$ is a chiral superfield \cite{FG}. 
Proving the integrability of an equation by linearization has been largely studied in the classical case \cite{GRL,RGT} and has found new developments in the $\mathcal{N}=1$ formalism \cite{CRG1}.
 We propose a similar development in the $\mathcal{N}=2$ formalism. In $\mathcal{N}=2$ SUSY, we consider a pair of supercovariant derivatives defined as
\begin{equation}
 D_1=\partial_{\theta_1}+\theta_1\partial_x,\quad D_2=\partial_{\theta_2}+\theta_2\partial_x,\label{covder}
\end{equation}
which satisfy the anticommutation relations $\{D_1,D_1\}=\{D_2,D_2\}=2\partial_x$ and $\{D_1,D_2\}=0$. We consider also the complex supercovariant derivatives
\begin{equation}
 D_{\pm}=\frac{1}{2}(D_1\pm i D_2),\label{complexder}
\end{equation}
which satisfy $\{D_{\pm},D_{\pm}\}=0$ and $\{D_+,D_-\}=\partial_x$. In terms of the complex Grassmann variables $\theta_{\pm}=\frac{1}{\sqrt{2}}(\theta_1\pm i\theta_2)$, the derivatives (\ref{complexder}) admits the following representation
\begin{equation}
 D_{\pm}=\frac{1}{\sqrt{2}}(\partial_{\theta_{\mp}}+\theta_{\pm}\partial_x)
\end{equation}
and the superfield $A$ given in (\ref{fieldA}) writes 
\begin{equation}
 A(x,t;\theta_+,\theta_-)=u(x,t)+\theta_+\rho_-(x,t)+\theta_-\rho_+(x,t)+i\theta_+\theta_-v(x,t).\label{chisuper}
\end{equation}
The fermionic complex valued functions $\rho_\pm$ are defined as $\rho_\pm=\frac{1}{\sqrt{2}}(\xi_1\pm i\xi_2)$.
% The supercovariant derivatives $D_{\pm}$ acts on $A$ in the following way
%\begin{equation}
% D_{\pm}A=\frac{1}{\sqrt{2}}\left(\rho_{\pm}+\theta_{\pm}(u_x\mp iv)\pm\theta_{+}\theta_{-}\rho_{\pm,x}\right).
%\end{equation}

Chiral superfields are superfields of type (\ref{chisuper}) satisfying $D_+A=0$. In terms of components, we get
\begin{equation}
 A(x,t;\theta_+,\theta_-)=u(x,t)+\theta_+\rho_-(x,t)+\theta_+\theta_-u_x(x,t),\label{chi}
\end{equation}
or equivalently $\xi_2=i\xi_1$ and $v=-iu_x$.

In the subsequent sections, we produce solutions of $\mathcal{N}=2$ SUSY extensions of the Korteweg-de Vries\cite{LM} (SKdV$_a$), modified Korteweg-de Vries\cite{ZP} (SmKdV) and Burgers\cite{HK} (SB) equations from a chiral superfield point of view.
In this instance, the equations, in terms of the complex covariant derivatives (\ref{complexder}), reads, respectively, as
\begin{align}
 A_t&=(-A_{xx}+i(a+2)A[D_+,D_-]A+i(a-1)[D_+A,D_-A]+aA^3)_x,\label{SKdV}\\
A_t&=-A_{xxx}-2A_x^3-6([D_+,D_-]A)^2A_x,\label{SmKdV}\\
A_t&=(i[D_+,D_-]A+2A^2)_x,\label{SB}
\end{align}
where $[X,Y]=XY-YX$ is the commutator. In (\ref{SKdV}), $a$ is an arbitrary parameter but we will consider only the integrable cases \cite{LM}  where $a=-2,1,4$.
 
In this paper, we start by presenting a general reduction procedure of these equations using chiral superfields (Section II). We thus treat SKdV$_{-2}$ and SmKdV together and construct classical $N$ super soliton solutions \cite{GS,DH,MY,ZLW} and an infinite set of similarity solutions \cite{DH}. In Section IV, we demonstrate the existence of special $N$ super soliton solutions, called virtual solitons \cite{HK}, for the SUSY extensions of the KdV equation with $a=1,4$ and the Burgers equation using a related linear partial differential equation.
The last section is devoted to a $\mathcal{N}=4$ extension of the KdV equation \cite{ZP} in an attempt to construct a general $N$ super virtual soliton solution.

%##################################### section ############################################

\section{General approach and chiral solutions.}

Here, we propose a general approach for the construction of chiral solutions of SUSY extensions. This approach avoids treating SUSY extensions in terms of components of the bosonic field $A$ given in (\ref{fieldA}). Assuming $D_+A=0$, we get the chiral property $\{D_+,D_-\}A=D_+D_-A=A_x$ and the equations (\ref{SKdV}), (\ref{SmKdV}) and (\ref{SB}) reduce to
\begin{align}
 A_t+(A_{xx}-i(a+2)AA_x-aA^3)_x&=0,\label{N2KdV}\\
A_t+A_{xxx}+8A_x^3&=0,\label{N2mKdV}\\
A_t-(iA_{x}+2A^2)_x&=0.\label{N2B}
\end{align}
Note that these equations may be evidently treated as classical\cite{DJ} PDE's, but remains SUSY extensions due to the grassmannian dependence of the bosonic field $A$. 

%Interestingly, equation (\ref{N2mKdV}) is obtained from equation (\ref{N2KdV}) via a spatial integration for $a=-2$. Thus, in the next section, we treat the SKdV$_{a=-2}$ and the SmKdV equations as one.

The absence of the grassmannian variables $\theta_+$ and $\theta_-$ derivatives in equations (\ref{N2KdV}), (\ref{N2mKdV}) and (\ref{N2B}), indicates that the odd sectors of chiral solutions
should be free from fermionic constraint. This property is in accordance with the integrability of these extensions due to arbitrary bosonization of the fermionic components\cite{GL} of the bosonic superfield $A$.

From the classical case, we know that the methods of resolution of all these equations are similar. The same could be said for the SUSY case. Indeed, if we assume the introduction of a potential $\tilde{A}$ such that $A= \tilde{A}_x$ in (\ref{N2KdV}) and after one integration with respect to $x$, we get
\begin{equation}
 {\tilde A}_t+{\tilde A}_{xxx}-i(a+2){\tilde A}_x{\tilde A}_{xx}-a {\tilde A}_x^3=0,\label{SKdV1}
\end{equation}
where the constant of integration is set to zero. The same is done on equation (\ref{N2B}) and leads to 
\begin{equation}
 {\tilde A}_t-i{\tilde A}_{xx}-2{\tilde A}_x^2=0,\label{SN2B}
\end{equation}
 We thus observe that the equations (\ref{N2mKdV}), (\ref{SKdV1}) and (\ref{SN2B}) are now on an equal footing, i.e.  the order of the equation in $x$ is equal to the number of appearance of $\partial_x$ in the nonlinear terms. This is standard in Hirota formalism.
The choice $a=-2$ in (\ref{SKdV1}) gives, up to a slight change of variable, the SmKdV equation (\ref{N2mKdV}). This means that the known\cite{DH} $N$ super soliton solutions and similarity solutions of SKdV$_{-2}$ will lead to similar types of solutions for the SmKdV equation (\ref{N2mKdV}). 

Now setting
\begin{equation}
 {\tilde A}(x,t;\theta_+,\theta_-)=\beta_{a}\ \log H_{a}(x,t;\theta_+,\theta_-)\label{boschan}
\end{equation}
in equation (\ref{SKdV1}), we obtain
\begin{equation}
H_{a}^2(H_{a,t}+H_{a,xxx})-(3+i(2+a)\beta_{a})H_{a}H_{a,x}H_{a,xx}- (\beta_{a}-i)(a \beta_{a}-2i)H_{a,x}^3=0.\label{linearcond}
\end{equation}
The above equation reduces to the linear dispersive equation \cite{DJ}
\begin{equation}
 H_{a,t}+H_{a,xxx}=0,\label{lineardis}
\end{equation}
for the special and only values $a=1$ with $\beta_1=i$ and $a=4$ with $\beta_4=\frac{i}{2}$. 
For $a=-2$, equation (\ref{linearcond}) writes
\begin{equation}
H_{-2}^2(H_{-2,t}+H_{-2,xxx})-3 H_{-2}H_{-2,x}H_{-2,xx}+2 (\beta_{-2}^2+1) H_{-2,x}^3=0,\label{linearcondmoins2}
\end{equation}
which does not linearize but can be bilinearized taking $\beta_{-2}=i$. It is discussed in the next Section.

A similar change of variable as in (\ref{boschan}) but with ${\tilde A}=\beta_B \, \log H_{B}$ and $\beta_{B}=\frac{i}{2}$ in (\ref{SN2B}) is assumed and leads to the linear Schr\"odinger equation
\begin{equation}
 H_{B,t}-iH_{B,xx}=0. \label{Schro}
\end{equation}

%###################################### section ##################################%

\section{SKdV$_{-2}$ and SmKdV equations}

It is well known\cite{MY,DJ,AS,C,FOU,GS1} that we can generate via the Hirota bilinear formalism $N$ soliton and similarity solutions in the classical case and 
in SUSY $\mathcal{N}=1$ extensions. Recently, the formalism was adapted to $\mathcal{N}=2$ extensions\cite{GS,DH,ZLW} by splitting the equation
into two $\mathcal{N}=1$ equations, one fermionic and one bosonic. Our approach consists of treating the equation as a $\mathcal{N}=2$ extension without splitting it, but imposing
chirality conditions.

Equation (\ref{N2mKdV}) can be bilinearized using the Hirota derivative defined as
\begin{equation}
 \mathcal{D}_x^n(f\cdotp g)=(\partial_{x_1}-\partial_{x_2})^nf(x_1)g(x_2)\vert_{x=x_1=x_2}.
\end{equation}
Indeed, we take ${\tilde A}$ as in (\ref{boschan}) with $\beta_{-2}=i$ and $H=\frac{\tau_1}{\tau_2}$, where $\tau_i=\tau_i(x,t;\theta_+,\theta_-)$ are bosonic chiral superfields for $i=1,2$. Equation (\ref{N2mKdV}) leads to the set of bilinear equations
\begin{align}
 (\mathcal{D}_t+\mathcal{D}_x^3)(\tau_1\cdotp\tau_2)&=0,\label{bili1}\\
\mathcal{D}_x^2(\tau_1\cdotp\tau_2)&=0\label{bili2}.
\end{align}
This set is analog to the corresponding bilinear equations in the classical mKdV equation\cite{DJ} but we deal with superfields  $\tau_1$ and $\tau_2$.

In order to get chiral solutions, we have to solve the set of bilinear equations
with the additional chiral property $D_+\tau_i=0$ for $i=1,2$. It will lead to new solutions of the SmKdV equation which are related to our recent contribution \cite{DH}.

%###############################subsubsection###########################%

\subsection{$N$ super soliton solutions}
The one soliton solution is easily retrieved. Indeed, we cast
\begin{equation}
 \tau_1=1+a_1\,e^{\Psi_1},\quad  \tau_2=1-a_1\,e^{\Psi_1},\label{tauone}
\end{equation}
where $a_1$ is an even parameter. $\Psi_1$ is a $\mathcal{N}=2$ chiral bosonic superfield defined as
\begin{equation}
 \Psi_1=\kappa_1 x-\kappa_{1}^3t+\theta_+\zeta_1+\theta_+\theta_-\kappa_1\label{exp}
\end{equation}
and never appears on this form in other approaches of $\mathcal{N}=2$  SUSY.
The parameters $\kappa_1$ and $\zeta_1$ are, respectively, even and odd. The $\tau$-functions (\ref{tauone}) together with (\ref{exp}) solve the set of bilinear equations (\ref{bili1}) and (\ref{bili2}) and give
rise to a one super soliton solution. Since $D_+\Psi_1=0$, the resulting traveling wave solution is chiral.
% For the two super soliton solution, we take
%\begin{align}
% \tau_1(x,t;\theta_+,\theta_-)&=1+a_1e^{\Psi_1}+a_2e^{\Psi_2}+a_1a_2A_{12}e^{\Psi_1+\Psi_2},\\
%\tau_2(x,t;\theta_+,\theta_-)&=1-a_1e^{\Psi_1}-a_2e^{\Psi_2}+a_1a_2A_{12}e^{\Psi_1+\Psi_2},
%\end{align}
%where $A_{12}=\left(\frac{\kappa_1-\kappa_2}{\kappa_1+\kappa_2}\right)^2$, $a_i$'s are non zero constants (for $i=1,2$) and $\Psi_i$'s are given as in (\ref{exp}).
 %Those $\tau$-functions clearly solves the bilinear equations (\ref{bili1}) and (\ref{bili2}) and makes the superfield $\tilde{A}=\frac{i}{2}\log\frac{\tau_1}{\tau_2}$ a chiral superfield.
% We get explicitly the components of the superfield solution $A$, of the SKdV$_{-2}$ equation, given by
%\begin{align}
% u(x,t)&=i\partial_x\log\left(\frac{1+a_1e^{\eta_1}+a_2e^{\eta_2}+a_1a_2A_{12}e^{\eta_1+\eta_2}}{1-a_1e^{\eta_1}-a_2e^{\eta_2}+a_1a_2A_{12}e^{\eta_1+\eta_2}}\right),\\
%\xi_1(x,t)&=2i\partial_x\left(\frac{a_1\zeta_1e^{\eta_1}+a_2\zeta_2e^{\eta_2}-a_1a_2A_{12}e^{\eta_1+\eta_2}(a_1\zeta_2e^{\eta_1}+a_2\zeta_1e^{\eta_2})}{1-a_1^2e^{2\eta_1}-a_2^2e^{2\eta_2}-2a_1a_2(1-A_{12})e^{\eta_1+\eta_2}+a_1^2a_2^2A_{12}^2e^{2(\eta_1+\eta_2)}}\right),
%\end{align}
%$v=-iu_x$ and $\xi_2=i\xi_1$, where $\eta_i=\kappa_ix-\kappa_i^3t$ for $i=1,2$. We may easily generalize the $\tau$-functions to obtain the so-called $N$ super soliton solutions \cite{GS,DH}.

Since, we exhibit the three super soliton solution of the SmKdV equation in figures 1 and 2, we give the general expressions of $\tau_1$ and $\tau_2$:
\begin{align}
 \tau_1(x,t;\theta_+,\theta_-)&=1+\sum_{i=1}^3a_ie^{\Psi_i}+\sum_{i<j}a_ia_jA_{ij}e^{\Psi_i+\Psi_j}+a_1a_2a_3A_{12}A_{13}A_{23}e^{\Psi_1+\Psi_2+\Psi_3},\label{3sol1}\\
\tau_2(x,t;\theta_+,\theta_-)&=1-\sum_{i=1}^3a_ie^{\Psi_i}+\sum_{i<j}a_ia_jA_{ij}e^{\Psi_i+\Psi_j}-a_1a_2a_3A_{12}A_{13}A_{23}e^{\Psi_1+\Psi_2+\Psi_3},\label{3sol2}
\end{align}
where $A_{ij}=\left(\frac{\kappa_i-\kappa_j}{\kappa_i+\kappa_j}\right)^2$ and the $\Psi_i$'s are defined as in (\ref{exp}). 
The functions $\tau_1$ and $\tau_2$ solves the bilinear equations (\ref{bili1}) and (\ref{bili2}) and are such that $D_+\tau_i=0$ for $i=1,2$. 
The generalization to a $N$ super soliton solution is direct using the $\tau$-functions expressed above. 
The forms of the $\tau$-functions given above are new representations of super soliton solutions and have never been introduced before.

In figure 1, we may enjoy the three soliton solution Im($v$) of the SmKdV equation given by
\begin{equation}
 v(x,t)=\frac12\partial_x\log\left(\frac{\tau_1(x,t;0,0)}{\tau_2(x,t;0,0)}\right),
\end{equation}
as a function of $x$, for the special values $\kappa_1=\frac43\kappa_2=2\kappa_3=\frac43$, $a_i=i$ in (\ref{3sol1}) and (\ref{3sol2}) and $t=-20,0,20$. In figure 2, we explore the behavior of the fermionic component
$\rho_-$ of the superfield $A$ for the same special values. To achieve this, we write $\rho_-$ as
\begin{equation}
 \rho_-(x,t)=\zeta_1\, f_1(x,t)+\zeta_2\, f_2(x,t)+\zeta_3\, f_3(x,t)
\end{equation}
and trace out the bosonic functions $f_1$, $f_2$ and $f_3$.

%In figure 1, we may enjoy the three soliton solution $-u$ of the SKdV$_{-2}$ equation for $\kappa_1=\frac43\kappa_2=2\kappa_3=\frac43$, $a_i=i$ in (\ref{3sol1}) and (\ref{3sol2}) and $t=-20,0,20$.
% In figure 2, we may enjoy the behavior of the function Im($v$) where $v=-iu_x$, $\kappa_1=\frac43\kappa_2=2\kappa_3=\frac43$, $a_i=i$ in (\ref{3sol1}) and (\ref{3sol2}) and $t=-20,0,20$.

\begin{figure}[!ht]
\begin{center}
\includegraphics[width=4cm]{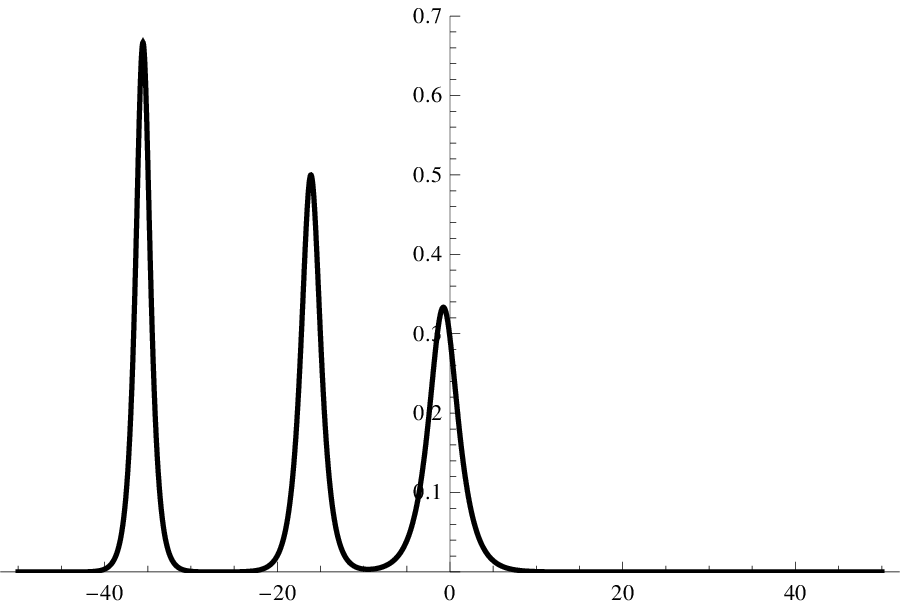}
\includegraphics[width=4cm]{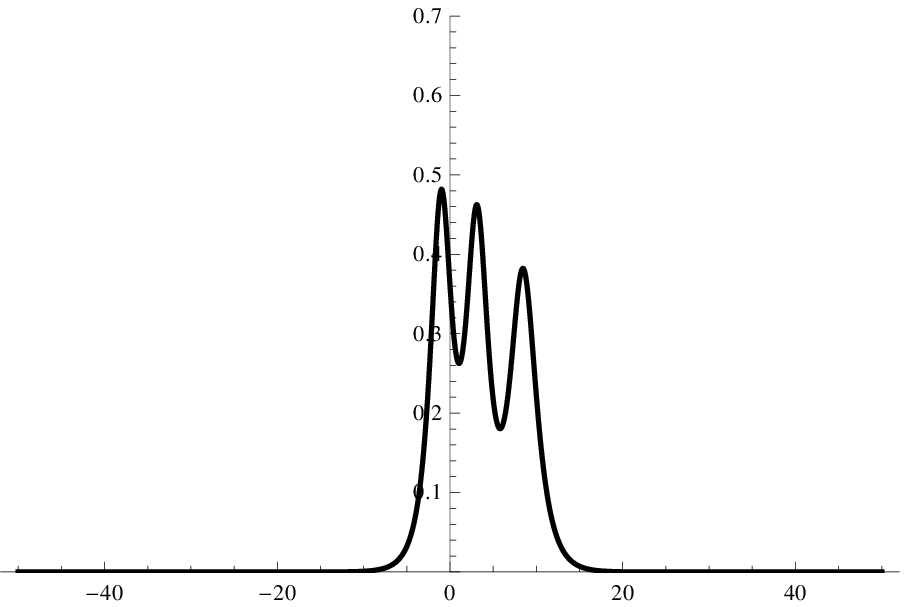}
\includegraphics[width=4cm]{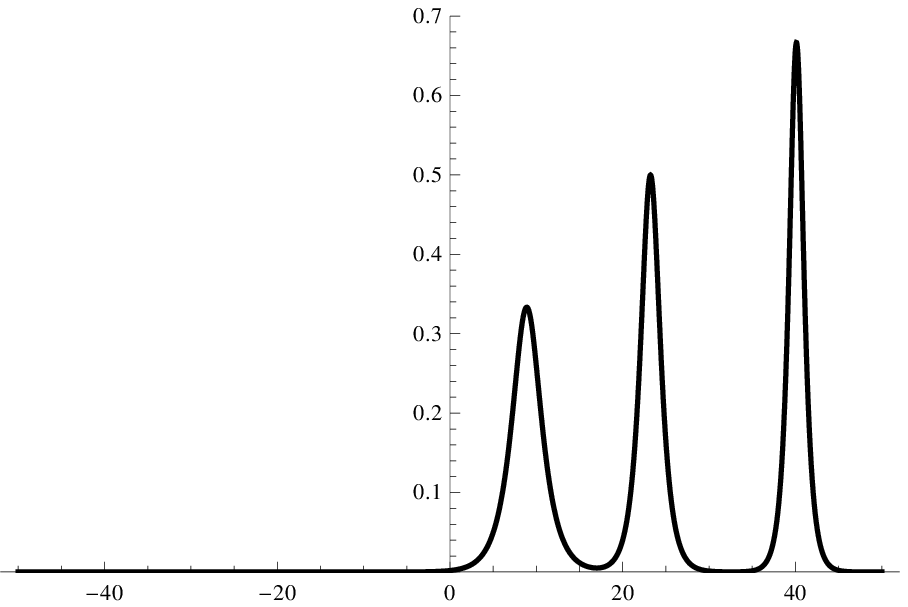}
\caption{The function Im($v$) of the three soliton solution of the SmKdV equation where $\kappa_1=\frac43\kappa_2=2\kappa_3=\frac43$  and $t=-20,0,20$.}
\end{center}
\end{figure}

\begin{figure}[!ht]
\begin{center}
\includegraphics[width=4cm]{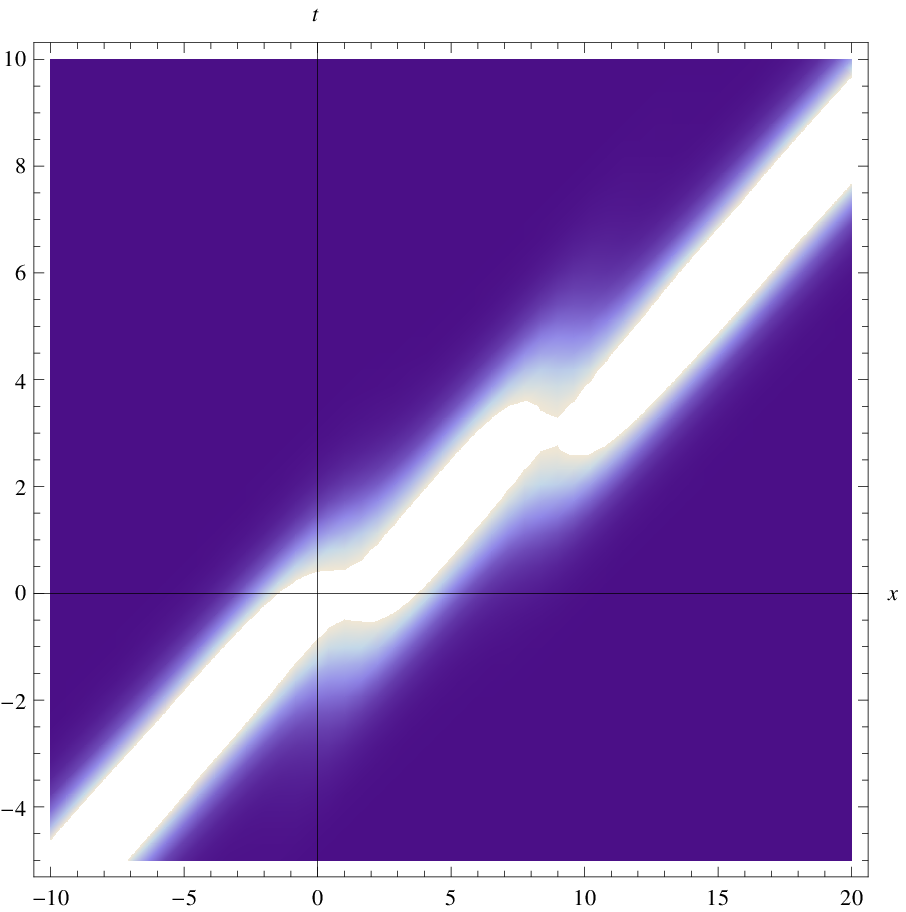}
\includegraphics[width=4cm]{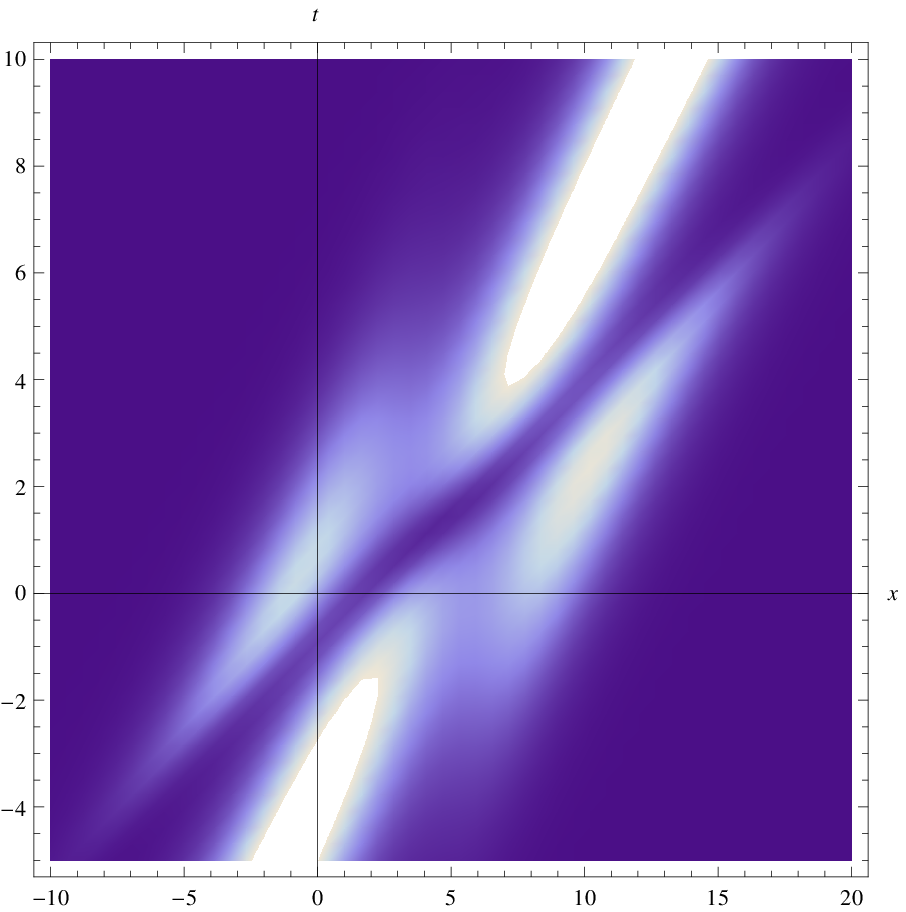}
\includegraphics[width=4cm]{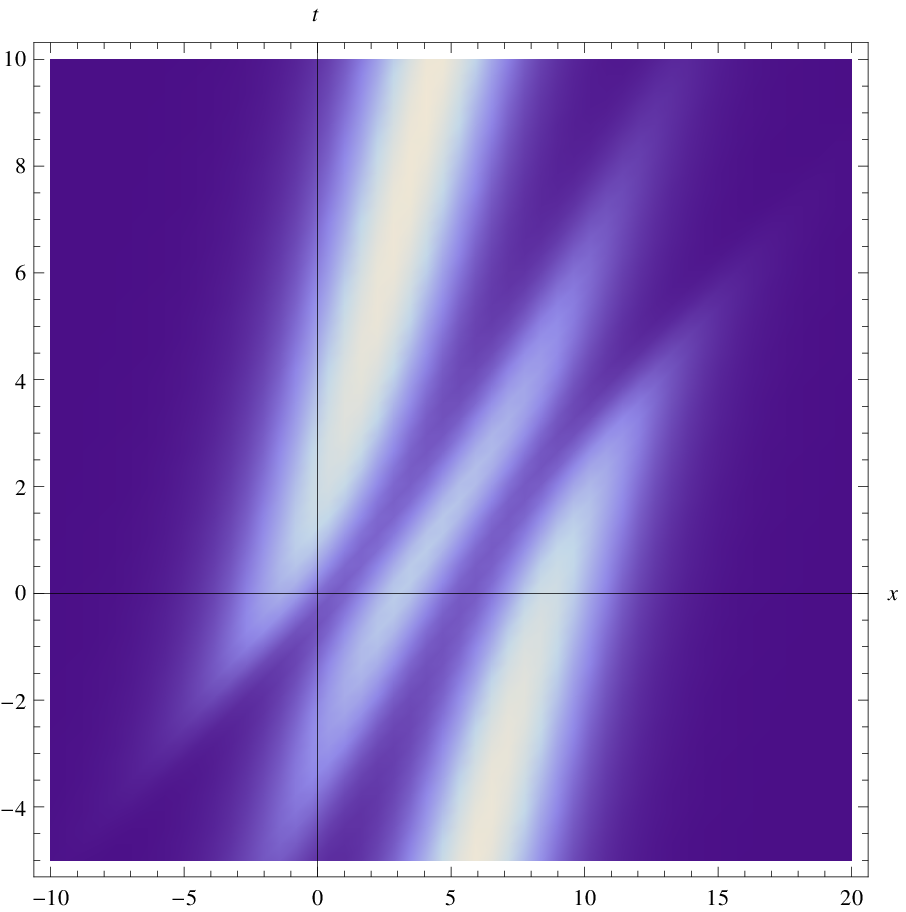}
\caption{The density plots of the functions $f_1$, $f_2$ and $f_3$, respectively from left to right, of the three soliton solution of the SmKdV equation where $\kappa_1=\frac43\kappa_2=2\kappa_3=\frac43$.}
\end{center}
\end{figure}

%#################################subsubsection###################################################%

\subsection{Similarity solutions}

In a recent paper \cite{DH}, we have proven the existence of an infinite set of rational similarity solutions of the SKdV$_{-2}$ using a SUSY version of the Yablonskii-Vorob'ev polynomials \cite{AS,C,FOU}.
 We propose in this subsection, to retrieve those solutions and find an infinite set of similarity solution for the SmKdV equation. 
To give us a hint into what change of variables we have to cast, we have used the symmetry reduction method associated to a dilatation invariance \cite{AHW}.

Let us define the following $\tau$-functions \cite{DH}
\begin{equation}
 \tau_{1,n}({\tilde z},t)=t^{\frac{n(n+1)}{6}}Q_n({\tilde z}),\label{tau1rat}
\end{equation}
where ${\tilde z}=t^{-\frac{1}{3}}(x+\theta_+\zeta+\theta_+\theta_-)$ and the functions $Q_n({\tilde z})$ are the Yablonskii-Vorob'ev polynomials defined by the recurrence relation
\begin{equation}
 3^{\frac{1}{3}}Q_{n+1}Q_{n-1}={\tilde z}Q_n^2-12(Q_nQ_{n,{\tilde z}{\tilde z}}-Q_{n,{\tilde z}}^2),
\end{equation}
with $Q_0({\tilde z})=3^{-\frac{1}{3}}$ and $Q_1({\tilde z})={\tilde z}$. We would like to insist that $\tilde{z}$ is a $\mathcal{N}=2$ bosonic superfield (as it is the case for the $\Psi_i$ in the preceding subsection).
 Using the fact that the Yablonskii-Vorob'ev polynomials satisfy the following bilinear equations \cite{C}
\begin{align}
 \left(\mathcal{D}_{{\tilde z}}^3-\frac{1}{3}{\tilde z}\mathcal{D}_{{\tilde z}}-\frac{n+1}{3}\right)(Q_n\cdot Q_{n+1})&=0,\\
\mathcal{D}_{{\tilde z}}^2(Q_n\cdot Q_{n+1})&=0,
\end{align}
we have that the pair of bilinear equations (\ref{bili1}) and (\ref{bili2}) are such that \cite{DH,AS,C,FOU}
\begin{align}
 (\mathcal{D}_t+\mathcal{D}_x^3)(\tau_{1,n}\cdot\tau_{1,n+1})&=0,\\
\mathcal{D}_x^2(\tau_{1,n}\cdot\tau_{1,n+1})&=0.
\end{align}
From the choice of the variable $\tilde{z}$, we also have $D_+\tau_{i,n}=0$ for all integers $n$. Taking $\tau_{2,n}=\tau_{1,n+1}$, we have an infinite set of similarity solutions of the SmKdV equation
given by
\begin{equation}
 \tilde{A}_n(\tilde{z},t)=\frac{i}{2}\log\left(\frac{\tau_{1,n}(\tilde{z},t)}{\tau_{1,n+1}(\tilde{z},t)}\right),
\end{equation}
for all integers $n\geq0$ and $\tau_{1,n}$ defined as in (\ref{tau1rat}). To get similarity solutions $A_n$ of the SKdV$_{-2}$, we use the above solution with $A_n=2t^{-\frac13}\partial_{\tilde{z}}\tilde{A}_n$. Plots of some similarity solutions are given in our recent contribution \cite{DH}.
%########################################subsection######################################%

\section{SKdV$_1$, SKdV$_4$ and SB equations and virtual solitons}

In this section, we exhibit $N$ super soliton solutions, called $N$ super virtual solitons, for the three equations SKdV$_1$, SKdV$_4$ and SB. 
Virtual solitons are soliton-like solutions which exhibit no phase shifts in nonlinear interactions. In terms of classical $N$ soliton solutions \cite{IMM,GS,HK,DH,DJ,GS1,AS}, this is equivalent to say that the interaction coefficients $A_{ij}$ between soliton $i$ and soliton $j$ are zero, $\forall i\neq j$. 
They manifest as  traveling wave solutions for negative time $t\ll0$ and decrease spontaneously at time $t=0$ to split into a $N$ soliton profile which exhibit no phase shifts. It is often said, that the traveling wave solution was charged with $N-1$ soliton, called virtual solitons\cite{HK}.

Using the change of variable (\ref{boschan}) for the unknown bosonic field $\tilde A$, we have seen that the bosonic field $H_a$ must be a chiral superfield and solve the linear dispersive equation  (\ref{lineardis}) when $a=1$ and $a=4$. 
For the Burgers equation, the bosonic field $H_B$ had to be chiral and solves (\ref{Schro}).

It is easy to show that they admit the following solution
\begin{equation}
H(x,t;\theta_+,\theta_-)=1+\sum_{i=1}^Na_ie^{\Psi_i},\label{serieexp}
\end{equation}
where the bosonic superfields $\Psi_i$ are given as
\begin{equation}
 \Psi_i=\kappa_ix+\omega(\kappa_i)t+\theta_+\zeta_i+\theta_+\theta_-\kappa_i.\label{expo}
\end{equation}
The frequencies $\omega(\kappa_i)$ are such that $\omega(\kappa_i)=-\kappa_{i}^3$  for SKdV$_a$and $\omega(\kappa_i)=i \,\kappa_{i}^2$ for SB. It looks like a typical KdV type soliton solution where all the interaction coefficients $A_{ij}$ are set to zero.

We see that the virtual solitons solutions of the SKdV$_1$ and SKdV$_4$ equations are completely similar due to the form of $\tilde A$ which differs only by the constant value of $\beta_a$. The expression of the original bosonic field is obtained from 
\begin{equation}
 A=\beta\,\frac{H_x}{H},
\end{equation}
where $\beta=\beta_a$ for the SKdV$_a$ equation and $\beta=\beta_B$ for the SB equation. Thus, we can give the explicit forms of the superfield components $u$ and $\rho_-$. Indeed, we have
\begin{equation}
 u(x,t)=\beta\,\frac{\sum_{i=1}^{N}a_i\kappa_ie^{\eta_i}}{1+\sum_{i=1}^{N}a_ie^{\eta_i}},\quad \rho_-(x,t)=\beta\,\sum_{i=1}^{N}\zeta_i\,f_i(x,t),\label{compvirt}
\end{equation}
where $\eta_i=\kappa_ix+\omega(\kappa_i)t$ and the bosonic functions $f_i(x,t)$ are defined as
\begin{equation}
 f_i(x,t)=\frac{a_i\kappa_ie^{\eta_i}+\sum_{j=1}^{N}a_ia_j(\kappa_i-\kappa_j)e^{\eta_i+\eta_j}}{\left(1+\sum_{j=1}^Na_je^{\eta_j}\right)^2}.\label{fermicom}
\end{equation}

In figure 3, we may enjoy the three virtual soliton solution Im($u$) of the SKdV$_{1}$ equation for $\kappa_1=\frac43\kappa_2=2\kappa_3=\frac43$  and $a_i=1$ in (\ref{serieexp}) and $t=0,10,20$.
 In figure 4, we observe the behavior of the function $v$ where $v=-iu_x$, $\kappa_1=\frac43\kappa_2=2\kappa_3=\frac43$  and $a_i=1$ in (\ref{serieexp}) and $t=-20,0,20$. For the same special values,
figure 5 give the density plots of the bosonic functions $f_1$, $f_2$ and $f_3$ as given in (\ref{fermicom}).

\begin{figure}[!ht]
\begin{center}
\includegraphics[width=4cm]{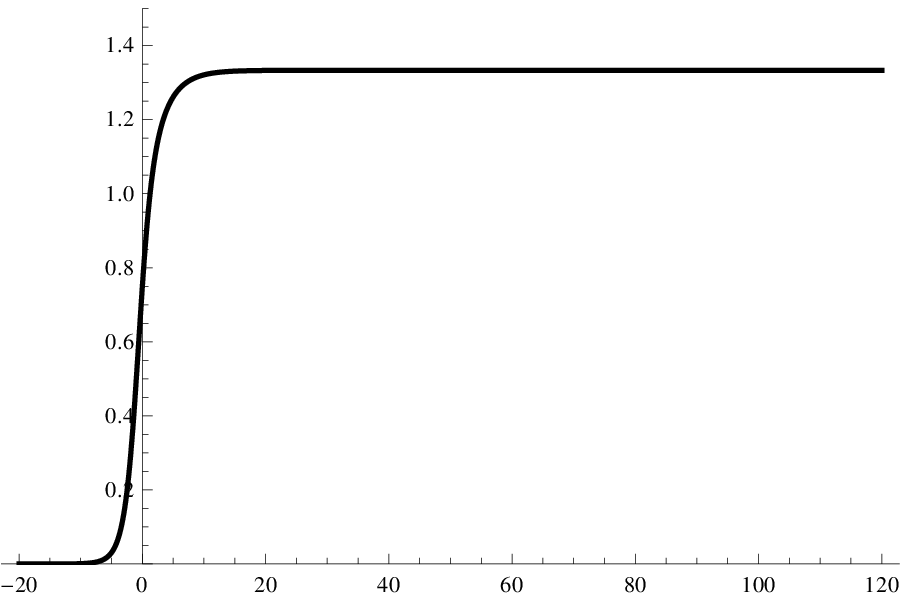}
\includegraphics[width=4cm]{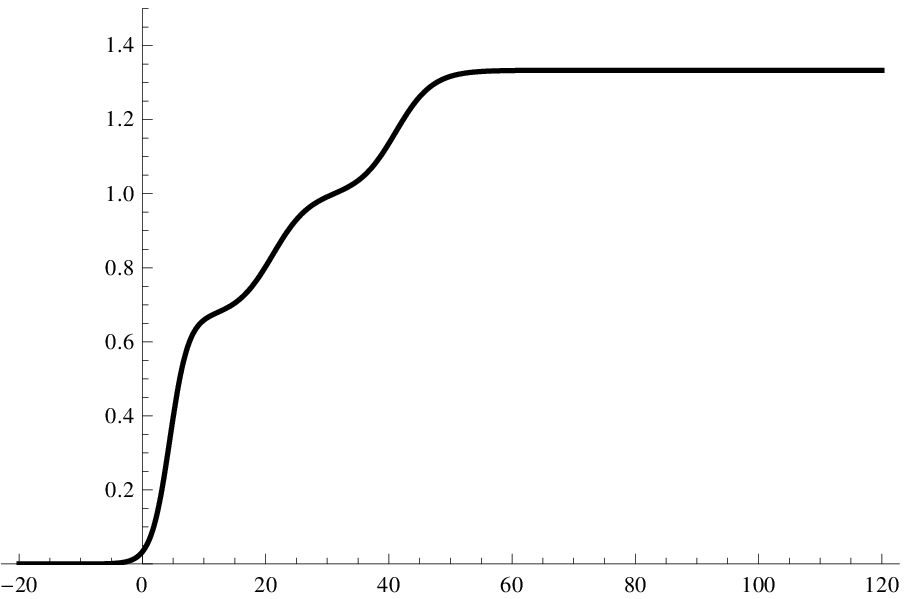}
\includegraphics[width=4cm]{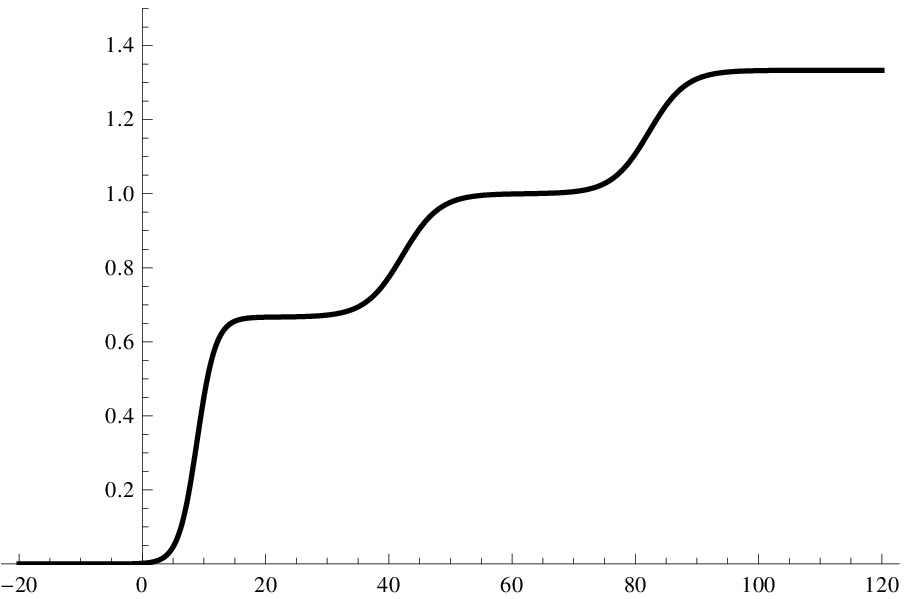}
\caption{The function Im($u$) of the three virtual soliton solution of the SKdV$_{1}$ equation where $\kappa_1=\frac43\kappa_2=2\kappa_3=\frac43$  and $t=0,10,20$.}
\end{center}
\end{figure}

\begin{figure}[!ht]
\begin{center}
\includegraphics[width=4cm]{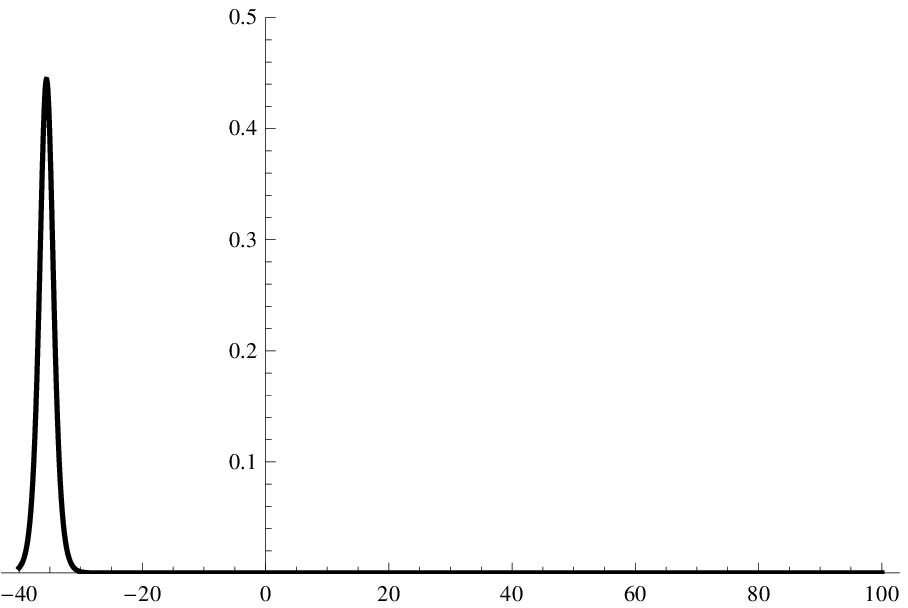}
\includegraphics[width=4cm]{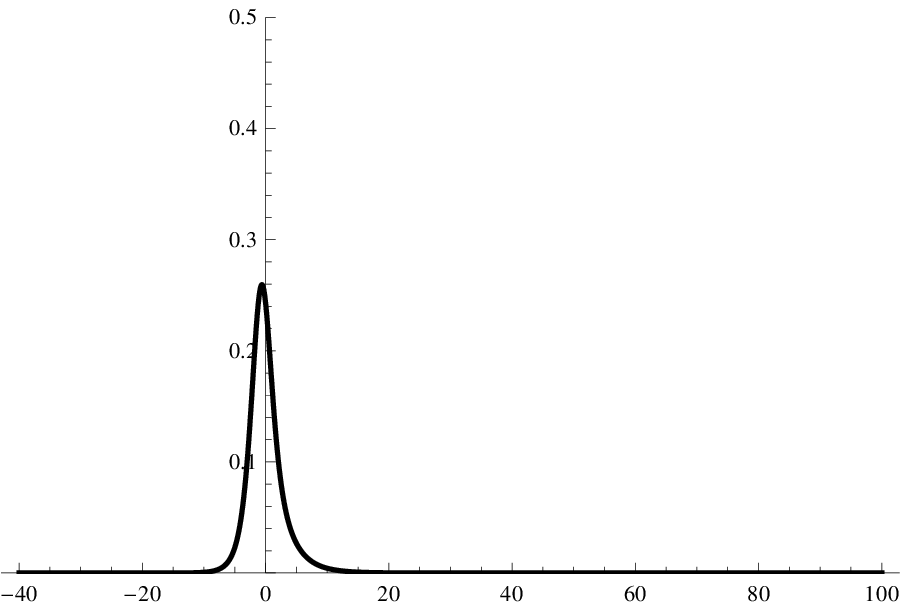}
\includegraphics[width=4cm]{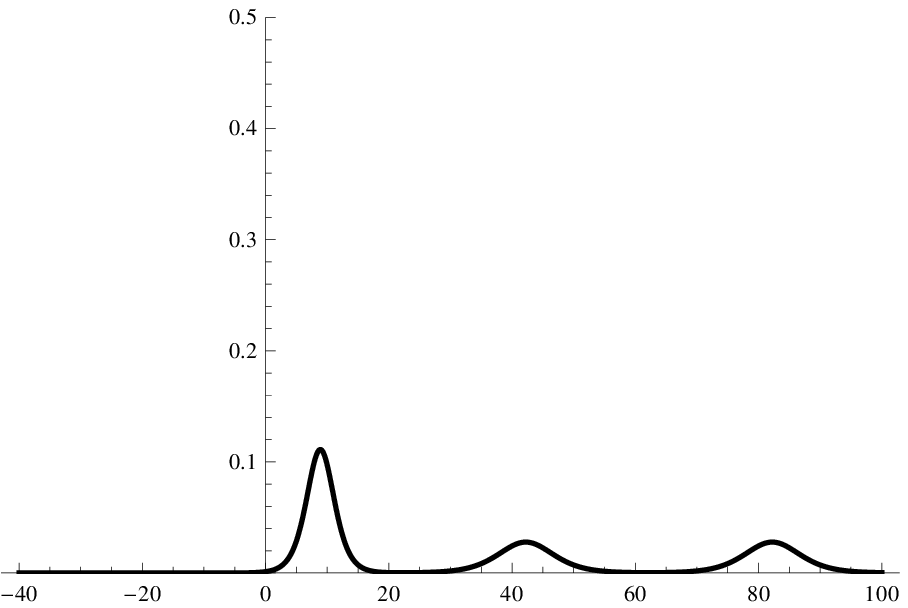}
\caption{The function $v$ of the three virtual soliton solution of the SKdV$_{1}$ equation where $\kappa_1=\frac43\kappa_2=2\kappa_3=\frac43$  and $t=-20,0,20$.}
\end{center}
\end{figure}

\begin{figure}[!ht]
\begin{center}
\includegraphics[width=4cm]{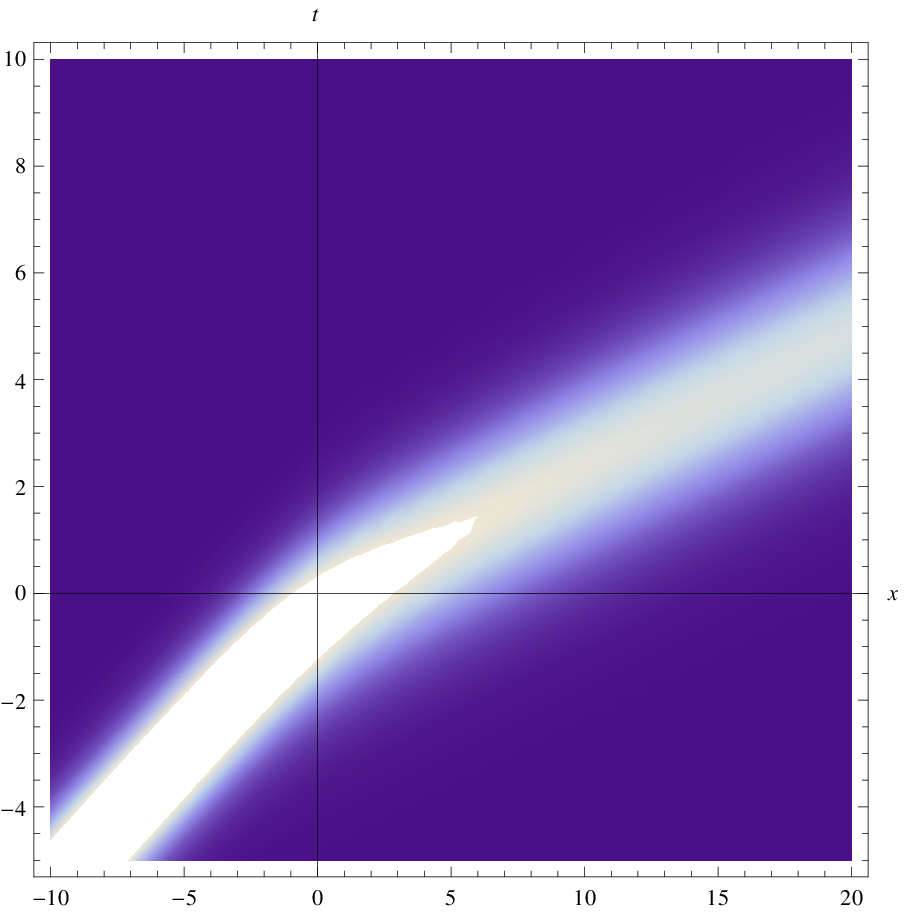}
\includegraphics[width=4cm]{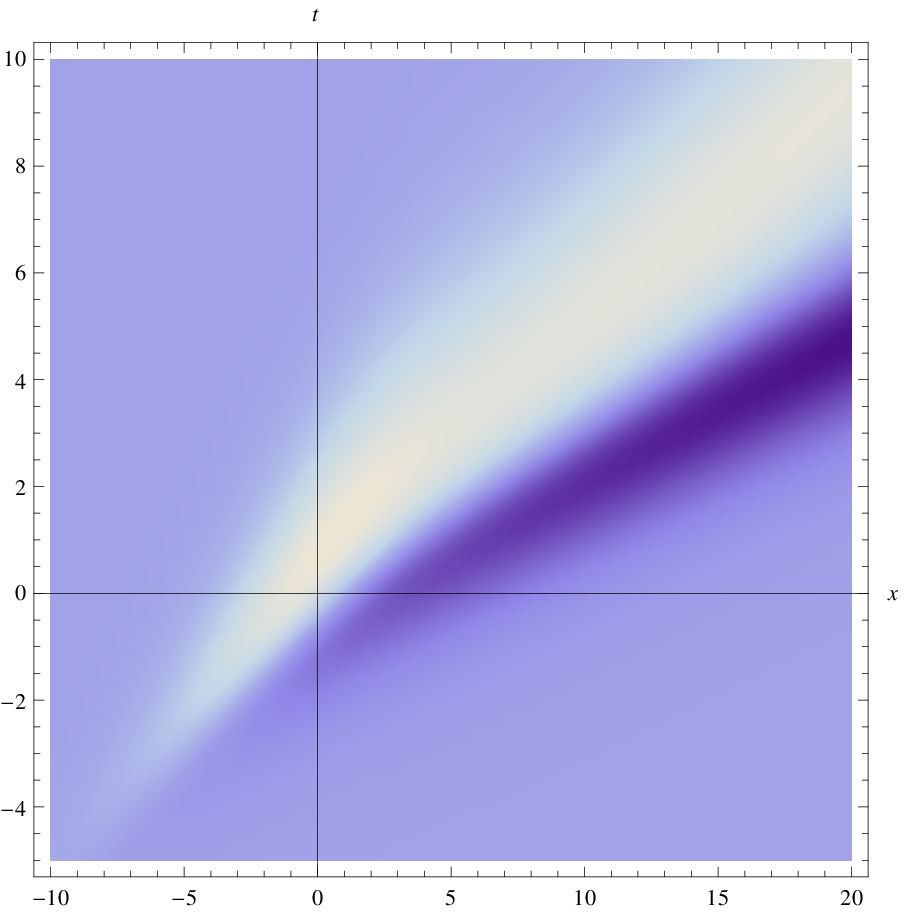}
\includegraphics[width=4cm]{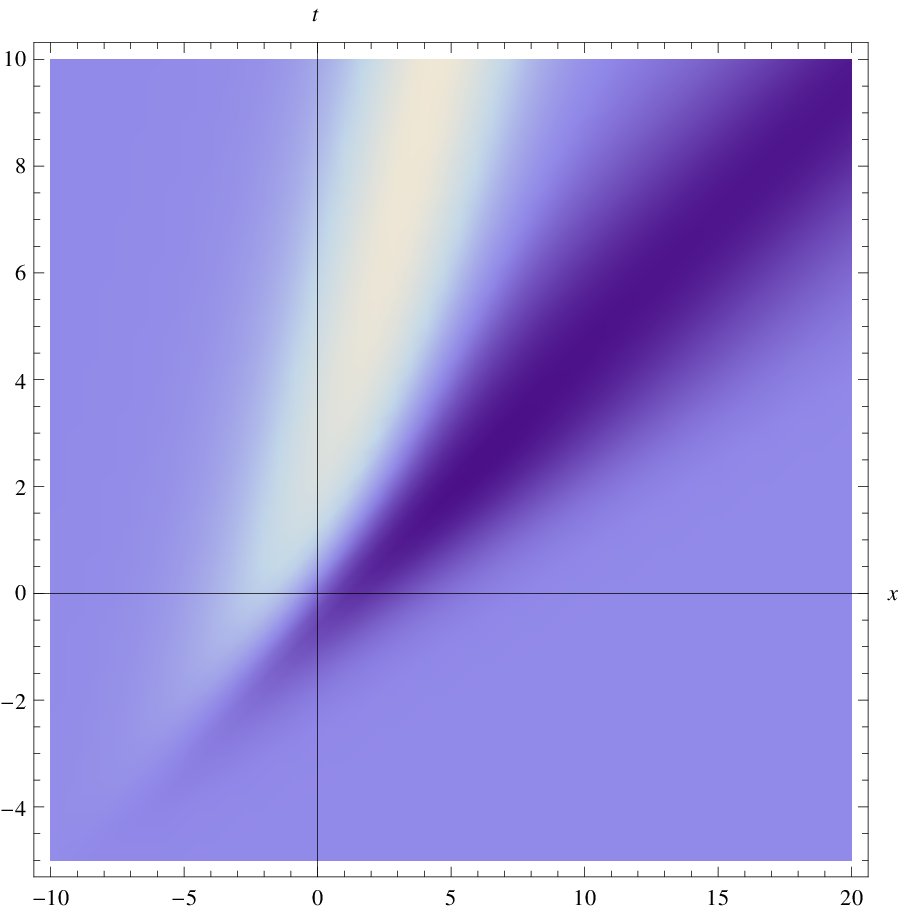}
\caption{The density plots of the functions $f_1$, $f_2$ and $f_3$, respectively from left to right, of the three virtual soliton solution of the SKdV$_1$ equation where $\kappa_1=\frac43\kappa_2=2\kappa_3=\frac43$.}
\end{center}
\end{figure}

%####################################subsection#######################################%

%\subsection{The SB equation and virtual solitons}

%As mention before the Burgers equation (\ref{N2B}) has been reduced to the linear Schr\"odinger equation (\ref{Schro}) through the change of variable $A= \beta_B \partial_x \log H_{B}$. The existence of $N$ super virtual solitons relies again on the linearity of (\ref{Schro}). In fact,
%\begin{equation}
% H(x,t;\theta_+,\theta_-)=1+\sum_{i=1}^{N}a_i\, e^{\Psi_i},
%\end{equation}
%solves equation (\ref{Schro}) for $a_i$ even parameters and the $\mathcal{N}=2$ chiral bosonic superfields $\Psi_j$ defined as 
%\begin{equation}
% \Psi_j=\kappa_jx+i\kappa_j^2t+\theta_+\zeta_j+\theta_+\theta_-\kappa_j,\quad j=1,...,N.
%\end{equation}

%To find the explicit forms of the components $u$ and $\rho_-$ of the bosonic superfield $A$, we use expression (\ref{compvirt}) where the $\eta_j$'s are replaced by $\eta_j=\kappa_jx+i\kappa_j^2t$.

%################################subsection##########################################

\section{SUSY $\mathcal{N}=4$ KdV equation and virtual solitons}

The SUSY $\mathcal{N}=4$ KdV equation, as proposed by Popowicz in \cite{ZP}, reads 
\begin{equation}
 \Gamma_t+\Gamma_{xxx}+4\Gamma_x^3+6[\check{D}_+,\check{D}_-](\Gamma_x[\hat{D}_+,\hat{D}_-]\Gamma)+12([\hat{D}_+,\hat{D}_-]\Gamma)^2\Gamma_x=0,\label{N4KdV}
\end{equation}
where $\Gamma$ is a bosonic superfield and the complex supercovariant derivatives are defined as
\begin{equation}
 \hat{D}_{\pm}=\frac{1}{2}(D_1\pm iD_2),\quad \check{D}_{\pm}=\frac{1}{2}(D_3\pm iD_4),\label{hcder}
\end{equation}
where $D_i=\partial_{\theta_i}+\theta_i\partial_x$ for $i=1,2,3,4$. 
Using the relations $\{D_i,D_j\}=2\,\delta_{ij}\,\partial_x$, where $\delta_{ij}$ is the Kronecker delta, we have that the supercovariant derivatives (\ref{hcder}) satisfy the anticommutation rules
\begin{equation}
 \{\hat{D}_{\mu},\check{D}_{\nu}\}=0,\quad \{\hat{D}_{\mu},\hat{D}_{\nu}\}=\{\check{D}_{\mu},\check{D}_{\nu}\}=(1-\delta_{\mu\nu})\partial_x,
\end{equation}
where $\mu,\nu\in\{+,-\}$. Equation (\ref{N4KdV}) can easily be viewed as a generalization of a $\mathcal{N}=2$ equation. Indeed, setting $\theta_3=\theta_4=0$ and $\Gamma=\frac{1}{\sqrt{2}}A$ in equation (\ref{N4KdV}), we retrieve
the SmKdV equation (\ref{SmKdV}).
% We introduce complex Grassmann variables defined as $\hat{\theta}_{\pm}=\frac{1}{\sqrt{2}}(\theta_1\pm i\theta_2)$ and $\check{\theta}_{\pm}=\frac{1}{\sqrt{2}}(\theta_3\pm i\theta_4)$ and, in this representation, the derivatives (\ref{hcder}) take the form
%\begin{equation}
 %\hat{D}_{\pm}=\frac{1}{\sqrt{2}}(\partial_{\hat{\theta}_{\mp}}+\hat{\theta}_{\pm}\partial_x),\quad \check{D}_{\pm}=\frac{1}{\sqrt{2}}(\partial_{\check{\theta}_{\mp}}+\check{\theta}_{\pm}\partial_x).
%\end{equation}

To construct virtual solitons of $\mathcal{N}=2$ SUSY extensions, we have considered chiral superfields.
 Here, we propose a generalization of this concept.
 Indeed, we impose the following constraints on the superfield $\Gamma$
\begin{equation}
 \hat{D}_+\,\Gamma=0,\quad \check{D}_+\,\Gamma=0.\label{chicond}
\end{equation}
A bosonic superfield $\Xi$ satisfying the chiral conditions (\ref{chicond}) has the following general form
\begin{align}
 \Xi(x,t;\hat{\theta}_{\mu},\check{\theta}_{\mu})&=u+\hat{\theta}_+\,\xi+\check{\theta}_+\,\eta+\hat{\theta}_+\hat{\theta}_-\, u_x+\check{\theta}_+\check{\theta}_-\, u_x+\hat{\theta}_+\check{\theta}_+\, w\label{chiral}\\\notag&+\hat{\theta}_+\hat{\theta}_-\check{\theta}_+\,\eta_x+\hat{\theta}_+\check{\theta}_+\check{\theta}_-\,\xi_x+\hat{\theta}_-\hat{\theta}_+\check{\theta}_-\check{\theta}_+\,u_{xx},
\end{align}
where $u=u(x,t)$ and $w=w(x,t)$ are complex valued bosonic functions and $\xi=\xi(x,t)$ and $\eta=\eta(x,t)$ are complex valued fermionic functions. 
The Grassmann variables in (\ref{chiral}) are defined as $\hat{\theta}_{\pm}=\frac{1}{\sqrt{2}}(\theta_1\pm i\theta_2)$ and $\check{\theta}_{\pm}=\frac{1}{\sqrt{2}}(\theta_3\pm i\theta_4)$.
Now, using the chirality conditions (\ref{chicond}), we have $\hat{D}_+ \hat{D}_-\,\Gamma=\check{D}_+ \check{D}_-\,\Gamma=\Gamma_x$ and equation (\ref{N4KdV}) reduces to the classical nonlinear PDE
\begin{equation}
 \Gamma_t+\Gamma_{xxx}+12\Gamma_x\Gamma_{xx}+16\Gamma_x^3=0.\label{classN4}
\end{equation}
Equation (\ref{classN4}) is, up to a slight change of variable, similar to equation (\ref{SKdV1}) for the integrable cases $a=1,4$. Indeed, we retrieve equation (\ref{SKdV1}) for $a=1,4$ by casting $\Gamma=-\frac{i}{12}(a+2)\tilde{A}$
in equation (\ref{classN4}).

The above equation can be linearized into the linear dispersive equation (\ref{lineardis}) by the change of variable
\begin{equation}
 \Gamma(x,t;\hat{\theta}_{\mu},\check{\theta}_{\mu})=\frac{1}{4}\log\, \Upsilon(x,t;\hat{\theta}_{\mu},\check{\theta}_{\mu}).
\end{equation}
Thus to obtain solutions of equation (\ref{N4KdV}), the superfield $\Upsilon$ must satisfy the constraints
\begin{equation}
 \Upsilon_t+\Upsilon_{xxx}=0,\quad \hat{D}_+\Upsilon=\check{D}_+\Upsilon=0.
\end{equation}
A solution to this system is
\begin{equation}
 \Upsilon=1+e^{\Psi_1}=1+e^{\kappa_1 x-\kappa_1^3t+\varphi_1(\hat{\theta}_{\mu},\check{\theta}_{\mu})},
\end{equation}
where $\varphi$ is a $\mathcal{N}=4$ chiral bosonic superfield of the form
\begin{equation}
 \varphi_1=\hat{\theta}_+\,\hat{\zeta}_1+\check{\theta}_+\,\check{\zeta}_1+(\hat{\theta}_+\hat{\theta}_-+\check{\theta}_+\check{\theta}_-)\,\kappa_1+\hat{\theta}_+\check{\theta}_+\, \lambda_1,\label{N4super}
\end{equation}
with $\hat{\zeta}_1^2=\check{\zeta}_1^2=0$ and $\lambda_1$ is an even constants. 
This result can thus be generalized to give a $N$ super virtual soliton solution of the SUSY $\mathcal{N}=4$ KdV equation (\ref{N4KdV}) by taking
\begin{equation}
 \Upsilon=1+\sum_{i=1}^N\, e^{\kappa_i x-\kappa_i^3t+\varphi_i(\hat{\theta}_{\mu},\check{\theta}_{\mu})},
\end{equation}
where the superfields $\varphi_i$ are defined as in (\ref{N4super}) for $i=1,...,N$. 

 It is interesting to note that setting $\check{\theta}_+=0$ in (\ref{N4super}), ones recover the superfields (\ref{exp}).

\section{Concluding remarks and future outlook}

In this paper, we have studied special solutions of supersymmetric extensions of the Burgers, KdV and mKdV equations in a unified way and using a chirality of the superfield $A$. 

We have recovered interacting super soliton solutions (often called KdV type solitons) and an infinite set of rational similarity solutions. To produce such rational solutions, we have used an SUSY extension of the Yablonskii-Vorob'ev polynomials. We have introduce a new representation of the $\tau$-functions to solve the bilinear equations. These $\tau$-functions are $\mathcal{N}=2$ extensions of classical $\tau$-functions of the mKdV equation. Till now, in the literature, only $\mathcal{N}=1$ extensions of the $\tau$-functions were given.

We have shown the existence of non interacting super soliton solutions, called virtual solitons, for the Burgers and SKdV$_a$ ($a=1,4$). These special solutions are a direct generalization of the solutions obtained in a recent contribution\cite{HK} where $N$ super virtual solitons have been found by setting to zero the fermionic contributions $\xi_1$ and $\xi_2$ in the bosonic superfield $A$ given as in (\ref{fieldA}). We retrieve those solutions by setting $\zeta_i=0$ in the exponent terms (\ref{expo}). Thus the chirality property, exposed in this paper, has produced a non trivial fermionic sector for a $N$ super virtual soliton. Furthermore, to obtain such solutions
we have related the SUSY equations to linear PDE's showing the true origin of those special solutions. 

A $\mathcal{N}=4$ extension of the KdV equation has been shown to produce a $N$ super virtual soliton solution.
The study of $\mathcal{N}=4$ extensions is quite new to us and we hope in the future to produce a $N$ super soliton solutions
 with interaction terms.
\section*{Acknowledgments.}
L. Delisle acknowledges the support of a FQRNT doctoral research scholarship. V. Hussin acknowledges the support of research grants from NSERC of Canada.

\bibliographystyle{unsrt}% plain,unsrt,alpha

\end{document}